\documentclass[10pt,conference]{IEEEtran}
\usepackage[utf8]{inputenc}

\usepackage{balance}
\usepackage{soul}
\usepackage{xcolor}
\usepackage{xspace}
\usepackage{url}
\usepackage{graphicx}
\usepackage{listings,multicol}
\usepackage[export]{adjustbox}
\usepackage{amsmath}
\usepackage{amssymb}
\usepackage{textcomp}
\usepackage{breakurl} 
\usepackage{listings}
\usepackage{cleveref}
\usepackage{algorithm}
\usepackage{lipsum}
\usepackage[noend]{algpseudocode}
\usepackage{booktabs}
\usepackage{makecell}
\usepackage{enumitem}

\newcommand{\ag}{Aggregation Model\xspace}

\newlist{questions}{enumerate}{2}
\setlist[questions,1]{label=\bfseries RQ\arabic*.,ref=RQ\arabic*}
\setlist[questions,2]{label=(\alph*),ref=\thequestionsi(\alph*)}

\newcommand{\toolname}{Aggregation of Stack Trace Similarities \\ for Crash Report Deduplication\xspace}

\newcounter{observation}
\newcommand{\observation}[1]{\refstepcounter{observation}
	\begin{center}
		\framebox{
			\begin{minipage}{0.93\columnwidth}
				{\bf Answer to RQ\arabic{observation}:} \textit{#1}
			\end{minipage}
		}
	\end{center}
}

\usepackage{mathrsfs}

\usepackage{pifont}

\usepackage{mathtools}
\usepackage{amsmath}
\usepackage{amssymb}
\usepackage{nccmath}

\usepackage[noadjust]{cite}

\makeatletter
\newcommand{\linebreakand}{%
  \end{@IEEEauthorhalign}
  \hfill\mbox{}\par
  \mbox{}\hfill\begin{@IEEEauthorhalign}
}
\makeatother

\usepackage{tikz}
\usetikzlibrary{positioning}
\usetikzlibrary{shapes,arrows}
\usepackage{multirow}
\usepackage{xcolor}

\title{\toolname}
\author{
\IEEEauthorblockN{Nikolay Karasov}
\IEEEauthorblockA{
    \textit{HSE University}\\
    nikolay.karasoff@gmail.com
}
\and
\IEEEauthorblockN{Aleksandr Khvorov}
\IEEEauthorblockA{
    \textit{HSE University} \\
    \textit{JetBrains}\\
    aleksandr.khvorov@jetbrains.com
}
\and
\IEEEauthorblockN{Roman Vasiliev}
\IEEEauthorblockA{
    \textit{JetBrains}\\
    roman.vasiliev@jetbrains.com
}
\linebreakand
\IEEEauthorblockN{Yaroslav Golubev}
\IEEEauthorblockA{
    \textit{JetBrains Research}\\
    yaroslav.golubev@jetbrains.com
}
\and
\IEEEauthorblockN{Timofey Bryksin}
\IEEEauthorblockA{
    \textit{JetBrains Research} \\
    timofey.bryksin@jetbrains.com  
}
}

\begin{document}

\maketitle

\begin{abstract}

    The automatic collection of stack traces in bug tracking systems is an integral part of many software projects and their maintenance.
    However, such reports often contain a lot of duplicates, and the problem of de-duplicating them into groups arises. 
    In this paper, we propose a new approach to solve the deduplication task and report on its use on the real-world data from JetBrains, a leading developer of IDEs and other software.
    Unlike most of the existing methods, which assign the incoming stack trace to a particular group in which a single most similar stack trace is located, we use the information about all the calculated similarities to the group, as well as the information about the timestamp of the stack traces.
    This approach to aggregating all available information shows significantly better results compared to existing solutions.
    The aggregation improved the results over the state-of-the-art solutions by 15 percentage points in the Recall Rate Top-1 metric on the existing NetBeans dataset and by 8 percentage points on the JetBrains data.
    Additionally, we evaluated a simpler $k$-Nearest Neighbors approach to aggregation and showed that it cannot reach the same levels of improvement.
    Finally, we studied what features from the aggregation contributed the most towards better quality to understand which of them to develop further.
    We publish the implementation of the suggested approach, and will release the newly collected industrial dataset upon acceptance to facilitate further research in the area.
\end{abstract}

\section{Introduction}\label{sec:introduction}

\begin{figure}[t]
\centering
    \includegraphics[width=\columnwidth]{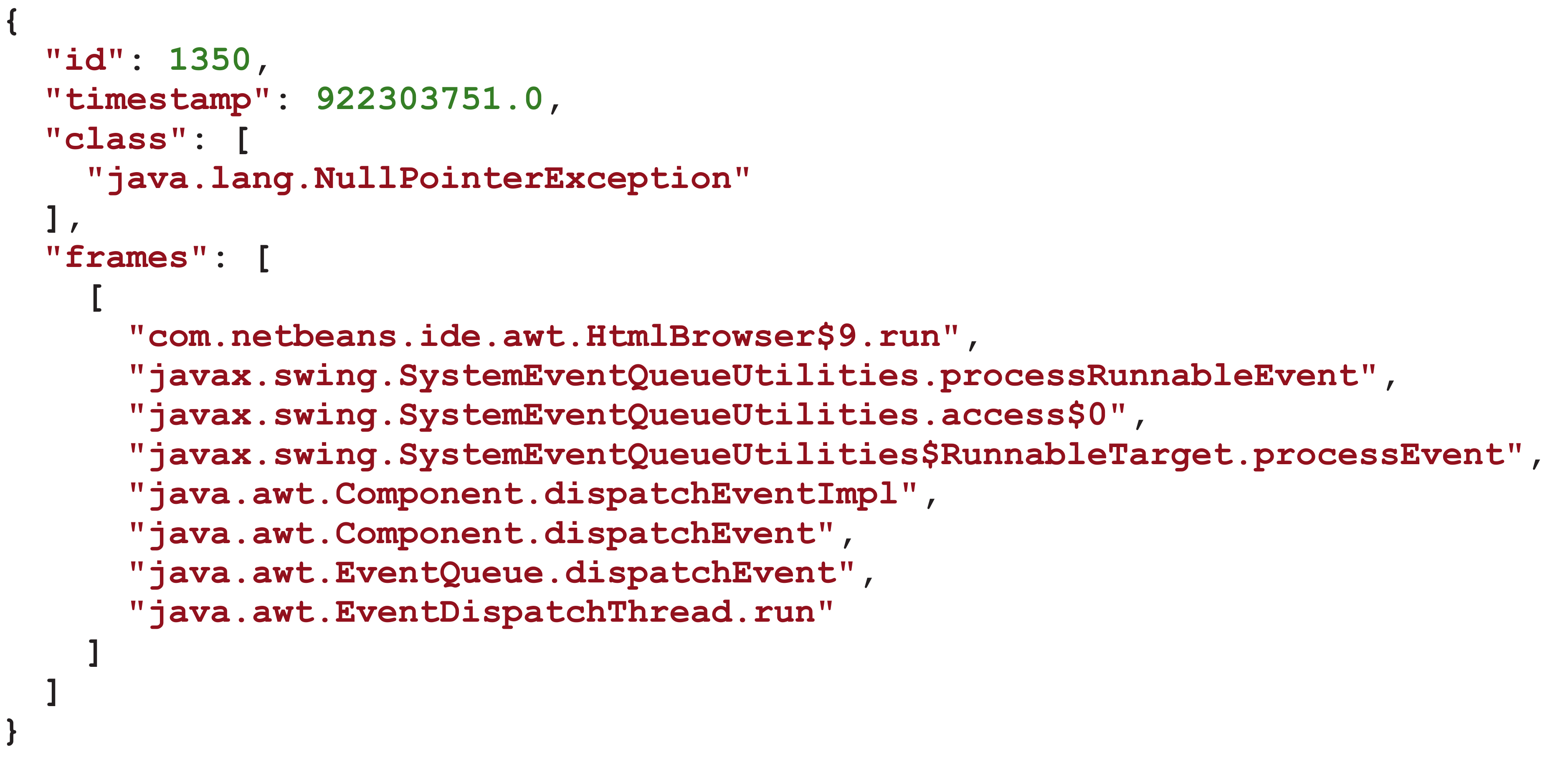}
    \centering
    \vspace{-0.4cm}
    \caption{An example of a stack trace from the NetBeans dataset~\cite{s3m}.}
    \vspace{-0.4cm}
    \label{fig:stacktrace}
\end{figure}

Software systems often use built-in bug tracking tools or crash reporting systems~\cite{crs_1,crs_2,crs_3} to get the information about the occurring crashes and errors from their users.
An important feature of such reports is that they often do not have a textual description of the problem, and the stack trace remains as the main source of information~\cite{s3m}.
An example of a stack trace is presented in~\Cref{fig:stacktrace}.
The stack trace consists of an ID, a timestamp, an error, and a list of frames --- an ordered list of methods that constitute the stack of calls that lead to the error.

For the exact same error, several reports can be recorded that become fuzzy duplicates~\cite{lerch,tracesim,s3m}.
Duplicates are caused by multiple users having a problem caused by the same error.
For example, $72\%$ of crash reports received for the IntelliJ Platform (a platform for building integrated development environments developed by JetBrains) are duplicates~\cite{tracesim}. 
Allowing developers to efficiently work with the reports requires us to de-duplicate them and group them by error type.
Since there are usually a lot of reports~\cite{durfex,many_reports_1}, it is not feasible to group them manually, as this implies a detailed study of each crash report, which is too time-consuming~\cite{researh_about_capacity}. 

Most modern methods~\cite{modani,lerch,durfex,tracesim,s3m} assign the incoming stack trace to one group or another based on the choice of the most similar stack trace in the group: locate a stack trace which is the most similar to the new one, and put the new one into that group. 
The disadvantage of this approach is that despite having all the similarity values to all stack traces in the group, we only use the information about the nearest one, and the overall structure of the group is not taken into account.
Some techniques also use the information about the group structure.
For example, the authors of CrashGraphs~\cite{crash_graphs} represent the group and the incoming stack trace as a directed graph. 
However, in this case, the situation is somewhat opposite: the information about the nearest stack trace in the group is lost. 

The results of these works that study different methods---based on the nearest stack trace~\cite{modani,lerch,durfex,tracesim,s3m} and based on the group structure~\cite{crash_graphs}---lead to the idea that both of these sources of information are important. 
In this work, we aim to develop an approach that would take into account the advantages of all the existing methods and test it on real-world data from a large software company.

The idea behind our approach is to use one of the existing methods to find the similarity between the incoming stack trace and stack traces in the group, and then aggregate the received information (the values of the calculated similarity, the number of stack traces in the group, etc.).
Another improvement that we propose is to use the information about the time of occurrence (timestamps) of the stack traces.
This is important because it gives us an idea of how far in time the group's stack traces are relative to the incoming stack trace. 
By aggregating all this information, we build an \ag, which determines the similarity between the stack trace and the group, thus making it possible to further rank the similarity to groups and select the closest one.

The approach is implemented as a simple and straightforward linear model that calculates the weighted sum of the constructed features.
The architecture of our approach is designed specifically to be as quick and efficient as it could to be used in production.
To evaluate the increase in the performance that the \ag brings about, we test it on two different datasets: the open-source dataset from the NetBeans IDE collected in our previous work~\cite{s3m}, as well as a new industrial dataset collected within JetBrains, a large developer of IDEs and other software. 
Upon acceptance, we will release this dataset to facilitate further research in the area.
The \ag shows the increase in the Recall Rate Top-1 metric ($\mathrm{RR} @ 1$) of $15$ percentage points for the NetBeans dataset and of $8$ percentage points for the JetBrains dataset, compared to existing state-of-the-art solutions. 
Such a significant difference demonstrates the usefulness and the potential of the proposed approach.

Next, we compare our approach with another simple way of aggregating the information about the similarities --- namely, various $k$-NN-based approaches. 
However, our experiments show that, when compared to the baselines, these approaches only reach the increase of $4$ percentage points in the $\mathrm{RR} @ 1$ metric on the NetBeans dataset, and $2$ percentage points on the JetBrains dataset, indicating that the \ag and the temporal information lead us to better performance.

Finally, to better understand the performance of the \ag, we study the coefficients of the model with which it aggregates the features.
As can be expected, the most important feature is the similarity to the nearest stack trace, the feature used in the majority of existing methods.
However, the results also show that the histogram of the similarities to all the stack traces in the group that takes timestamps into account helps the model as well.
This analysis can help researchers improve the \ag in the future.

Overall, our contributions in this paper are as follows:
\begin{itemize}
    \item \textbf{\ag}: We propose a new approach that takes into account all the available stack traces and their time of occurrence. 
    The implementation of the model with documentation is available online on GitHub: \url{https://github.com/nkarasovd/AggregationModel}. 
    \item \textbf{Dataset}: We collected a new large dataset using the anonymized proprietary data from the JetBrains software company, it contains 236,174 stack traces and 9,163 groups. 
    We will release the dataset upon acceptance.  
    \item \textbf{Evaluation}: We describe our experience with the \ag, which experimentally shows its usefulness for stack trace grouping. 
    Our approach outperforms existing methods by $15$ and $8$ percentage points in Recall Rate Top-$1$ metric on the NetBeans and JetBrains datasets, respectively.
\end{itemize}
\section{Background}\label{sec:background}

\subsection{Grouping Stack Traces}

There are various approaches for grouping stack traces that use different techniques, from edit distance to deep learning methods.
In this section, we describe the key ones.
It should be noted right away that most methods assign the incoming stack trace to the group that contains the most similar stack trace and therefore such approaches mainly differ in the way of calculating the similarity between stack traces.

Modani et al.~\cite{modani} use algorithms on strings to calculate stack trace similarity.
The authors present three similarity measures: edit distance, Longest Common Subsequence (LCS), and prefix match.
These algorithms are variants of classical string matching algorithms, with some modifications for a specific task of stack trace similarity.
In addition, the authors propose approaches for identifying uninformative functions (\textit{e.g.}, redundant recursion) and their further removal.

Lerch and Mezini~\cite{lerch} propose to use the approach based on the term frequency and the inverse document frequency (TF-IDF)~\cite{tfidf}.
In their work, each stack trace is tokenized, and the similarity value for the incoming stack trace $q$ and the given stack trace $d$ is calculated according to the following formula:
\begin{gather*}
    \mathrm{similarity}(q, d) = \sum_{t \in q} \mathrm{tf}_d(t) \cdot \mathrm{idf}(t)^2.
\end{gather*}
The advantage of the proposed approach is the speed of learning and the ease of implementation.

Sabor et al.~\cite{durfex} developed DURFEX, a technique that substitutes function names with the names of the packages in which they are defined to calculate the stack trace similarity, and then segments the resulting stack traces into N-grams of variable length.
This approach significantly reduces the amount of vocabulary used. This technique works with the JVM languages.

In our previous work, we proposed TraceSim~\cite{tracesim}, an approach for determining the similarity between two stack traces, which combines TF-IDF and string distance.
To obtain the values of hyperparameters used in the calculation of local and global frame weights, we formulate an optimization problem and use machine learning approaches to solve it.

Later, building on the ideas of processing stack traces, we proposed a new approach called S3M~\cite{s3m}.
S3M uses a Siamese neural network with biLSTM~\cite{lstm, bilstm} encoder to build the embeddings of stack traces and determine their similarity.
As a classifier, two fully-connected layers with ReLU activation are used. This is one of the first works in this field that uses deep learning methods.

Unlike previous works, Kim et al.~\cite{crash_graphs} proposed an approach called CrashGraphs that differs from most modern solutions described above in that it finds the similarity directly between the incoming stack trace and the groups themselves.
To do this, they represent the stack trace and the group in the form of a graph.
Firstly, all stack traces in the group are divided into pairs of consecutive calls.
For example, the group of two stack traces $ABC$ and $ACD$ will be parsed into pairs of frames $A \rightarrow B$, $B \rightarrow C$, $A \rightarrow C$, and $C \rightarrow D$.
Then, an oriented graph is built from these edges, and the similarity of a stack trace to the group is calculated by finding their sub-graph similarity. 
The main advantage of this approach is that we immediately get the similarity between the stack trace and the group, avoiding counting the similarity between all the individual stack traces.
In addition, such a representation of the group allows us to take into account its various features: composition, size, and uniqueness of stack traces.

The methods considered earlier determine the similarity value between the incoming stack trace and a certain group of stack traces as the similarity value between the incoming stack trace and the most similar stack trace in this group.
The main disadvantage of this technique is that finding this one most similar stack trace still requires calculating \textit{all} the similarity values for all stack traces, and this information is then discarded.
At the same time, the calculated similarity values can become a new source of information, which can improve the quality of many algorithms.
In addition, none of the considered works uses the information about the time of occurrence of stack traces in any way, which can also become a new source of information for future methods.

In this work, we aim to combine the advantages of the proposed approaches, create an \ag based on them that uses the time of occurrence of stack traces, and evaluate its usefulness on the real-world data from JetBrains, a large software company.

\subsection{Aggregating the Information with $k$-NN}
\label{sec:knn}

Note that placing the incoming stack trace into the group that contains the most similar stack trace is similar to the approach of the $k$-Nearest Neighbors algorithm ($k$-NN) with $k=1$. Let us briefly recall the idea of the $k$-NN approach.

Let $D$ be a set of labeled objects with labels from a certain set $Y$.
For a new object $x$, it is necessary to determine whether it belongs to a particular class.
The choice of a class is carried out according to the following decision function:
\begin{gather*}
    h(x, D) = \arg\max_{y \in Y}\sum_{x_i\in D}[y_i = y]w(x_i, x),\\
    w(x_i, x) = \begin{cases}1, \text{if $x_i$ is in the $k$ nearest neighbours of $x$}\\ 0, \text{otherwise}\end{cases}.
\end{gather*}
In the classical version of $k$-NN, the incoming object is assigned to the class whose objects were the most present among the $k$ closest to the new object.
If we take $k$ equal to 1, then we get an approach similar to choosing the group that contains the most similar stack trace.

The advantage of methods based on the idea of $k$-NN is that they do not require the presence of a metric space, with the main requirement being the ability to determine the distance (in our case, similarity) between a pair of objects.
Thus, the idea of $k$-NN can be transferred to the problem of grouping stack traces. 
In this case, taking into account multiple closest stack traces among the available groups will mean considering not only the single closest stack trace, but also other similiarity values in the groups.

A common modification to a basic $k$-NN approach is the \textit{weighted} $k$-NN approach.
The idea is that each object from the $k$ nearest neighbours is included in the decision rule with a certain non-zero weight.
A common approach to determining weight is to use kernel functions.
A lot of different kernel functions have been used in research: uniform, triangle, epanechnikov~\cite{epanechnikov}, quartic, triweight, gaussian, cosine, tricube~\cite{tricube}, logistic, sigmoid, silverman~\cite{silverman}.
These functions weight objects differently depending on their distance.

In addition to classical approaches, there are works that offer new types of weighting objects based on the distance.
Hyukjun et al.~\cite{k_conditional} propose the approach which calculates the distance between a new instance and the $k$-th nearest neighbor from each class, estimates posterior probabilities of class memberships using the distances, and assigns the instance to the class with the largest posterior. 
Ekin et al.~\cite{distance_based} propose several methods for weighting distances.
For example, in the \textit{Adjusted Weighting Method}, the distance to a certain class $Y_j$ is defined as the sum of the distances to the $\gamma$ nearest neighbours in this class with some decreasing weight $w$:
\begin{gather*}
    h(x, Y_j) = \sum_{i=1}^{\gamma}w^i \cdot d_i
\end{gather*}

Overall, $k$-NN-based approaches allow us to take into account the similarity to all stack traces in the groups.
To make sure that the proposed \ag is necessary, and that the same performance cannot be obtained by simply using a $k$-NN-based approach, we evaluate them also.
\section{Motivating example}\label{sec:motivation}

\begin{figure}[t]
\centering
    \includegraphics[width=\columnwidth]{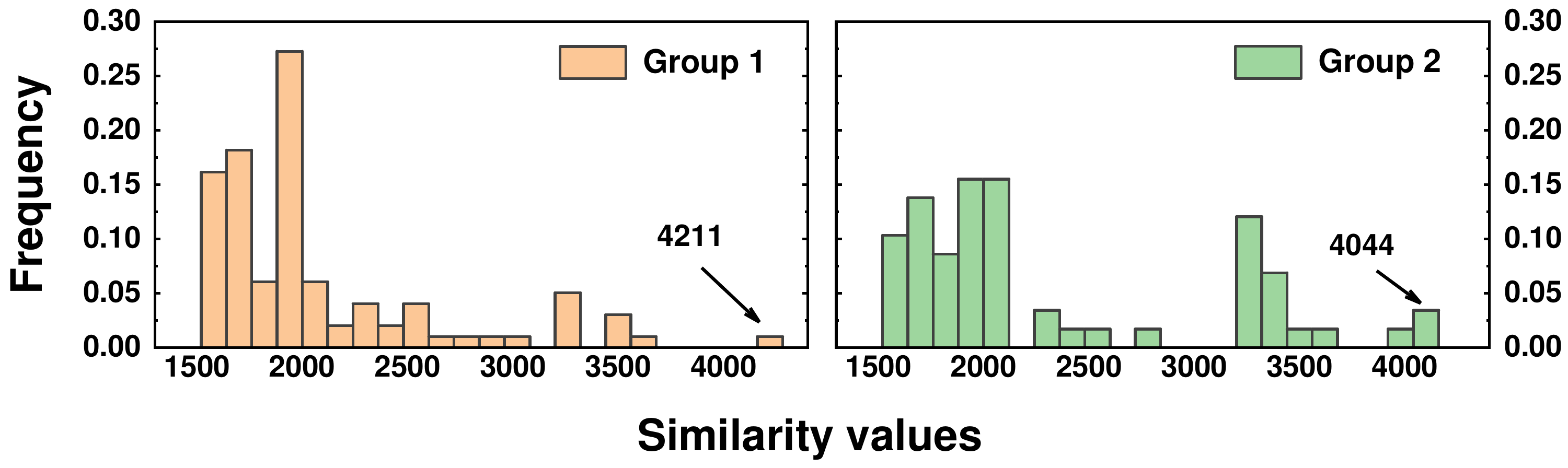}
    \centering
    \vspace{-0.4cm}
    \caption{An example distributions of similarity values of a certain incoming stack trace to stack traces of two groups.}
    \label{fig:hists}
\end{figure}

Consider the following example, presented in \Cref{fig:hists}.
This example was taken from the NetBeans data, a labeled dataset collected in our previous work~\cite{s3m}. 
Using the approach proposed by Lerch and Mezini~\cite{lerch}, we calculated the values of the similarity between a potential incoming stack trace and all stack traces in all groups, two of which are shown in the figure. 
It can be seen that the values of similarity differ among themselves very strongly inside each group.
For $Group_1$, the stack trace with the highest similarity to the incoming one (the right-most one) is very different from the rest of the group, indicating that it might not be suitable to make a judgement about the entire group.
The values of the maximum similarity for the $Group_1$ and $Group_2$ are equal to $4211$ and $4044$, respectively.
Thus, if selecting the group based on the most similar stack trace, the stack trace should be assigned to $Group_1$.
However, in this case, $Group_2$ is the \textit{ground truth} group, into which the stack trace should actually be defined based on the historical data.
This example clearly demonstrates the drawbacks of the most common way of assigning groups.

We believe that if we use the information about the values of all the similarities and the information about the structure of the group, as well as the time of occurrence of the stack traces, then this will avoid an error and correctly assign the incoming stack trace to $Group_2$.

\section{Approach}\label{sec:internals}

\begin{figure}[h!]
\centering
    \includegraphics[width=\columnwidth]{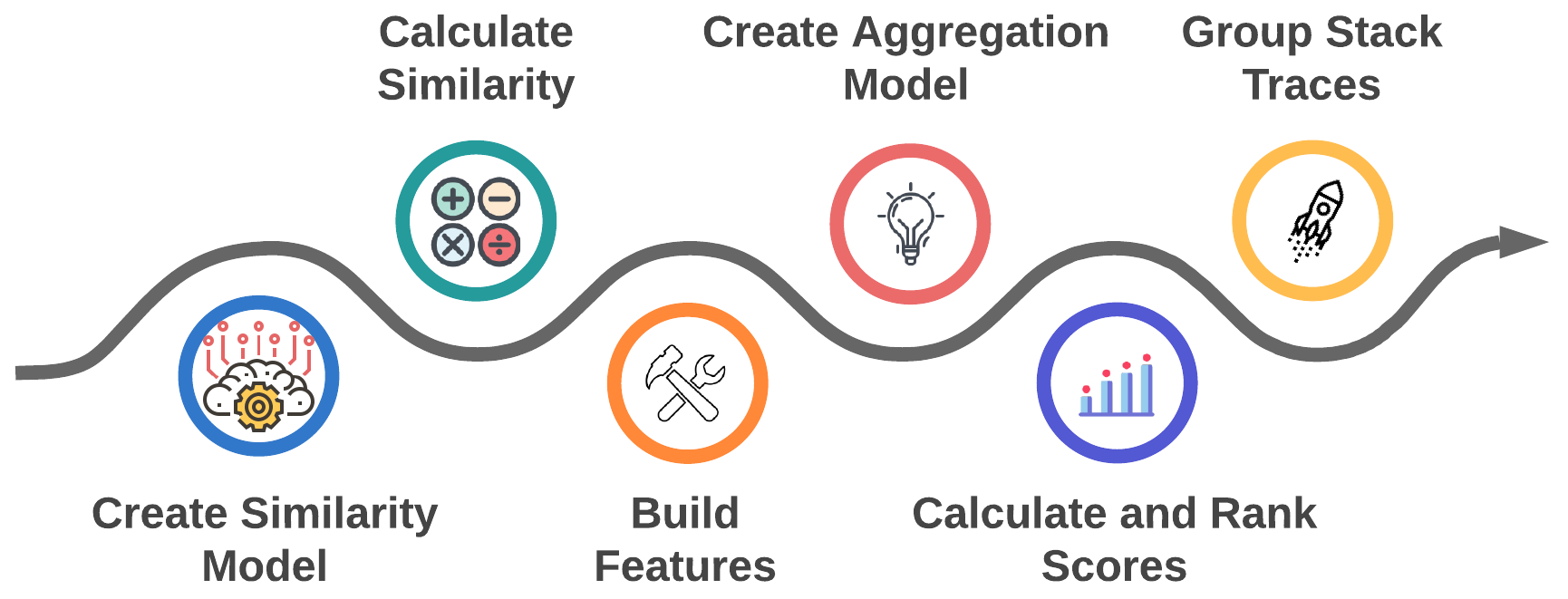}
    \centering
    \vspace{-0.4cm}
    \caption{A high-level pipeline of our approach.}
    \label{fig:scheme}
\end{figure}

\subsection{Overview}

An important feature of working with stack traces in large industrial projects is that the number of groups and their composition is constantly changing, new stack traces come in, new groups are formed. 
Therefore, we treated the task of assigning the new incoming stack trace to the best group as a ranking problem, and not a classification problem. 

The general pipeline of the proposed approach is presented in~\Cref{fig:scheme}.
The first step in our approach is to choose the similarity model for calculating the similarity between stack traces.
A model from any modern approach described above~\cite{modani,lerch,durfex,tracesim,s3m} can fit this role, making it possible for our technique to improve any of them.
Then, it is necessary to calculate the similarity between the incoming stack trace and all the stack traces of the group, for all groups.
The key difference between our approach and existing approaches is the \ag.

The \ag is a decision function that determines the similarity between the stack trace and the group.
This is done not based on the choice of the most similar stack trace, but by aggregating all the available information, which we transfer to it in the form of pre-calculated features.
We propose to build features based not only on the information about the values of similarity between stack traces, but also taking into account more complex information, such as the size of the group, its structure, as well as the time of occurrence of stack traces, etc.
The main motivation for using the time of occurrence is that the last added stack trace probably characterizes the group better than the ones that were added at the very beginning, since the group can change over time.

Finally, after the similarity between the incoming stack trace and all the groups is determined, we rank the groups according to the similarity and assign the stack trace to the most similar group. Let us now walk through the pipeline in greater detail.

\subsection{Preprocessing}
In order to calculate the similarity between stack traces, it is necessary to construct their vector representation. 
Initially, each stack trace 
is represented as a sequence of frames $S=\{f_1, f_2, \ldots, f_n\}$, where $f_i$ is the $i$-th frame. 
For each used similarity model, we apply the preprocessing as described in their work.
Most similarity models represent the stack trace as a sequence of tokens.
We take the full name of the method as its tokens, however, some models use the name of the class or the package.

\subsection{Features}\label{sec:features}

\begin{table*}[]
\centering
\caption{The description of features used in the \ag.}
\begin{tabular}{cp{4cm}p{12cm}}
\toprule

  \textbf{No.} &\multicolumn{1}{c}{\textbf{Feature name}} &
  \multicolumn{1}{c}{\textbf{Description}}  \\ 
  \midrule\midrule
  
  \multicolumn{3}{c}{\textbf{Features based on similarity}}\\\midrule\midrule

  \textbf{1} & First maximum & The value of similarity to the most similar stack trace in the group. This feature constitutes the base of the majority of methods and gives a very good ranking quality~\cite{lerch, durfex, s3m}.
  \\\midrule\midrule
  
  \multicolumn{3}{c}{\textbf{Features based on timestamps}}
  \\\midrule\midrule

  \textbf{2} & Maximum weight & The maximum weight corresponds to the stack trace in the group that is closest in time to the incoming one, which allows us to determine how long ago the group was updated relative to it. \\\midrule
  
  \textbf{3} & Minimum weight & The minimum weight corresponds to the stack trace in the group that is the farthest away in time from the incoming one, which allows us to determine how long ago the group was formed.
  \\\midrule
  
  \textbf{4} & Mean weight & The average value of all weights in the group. This shows us how far the incoming stack trace is from the group in time on average.
  \\\midrule
  
  \textbf{5} & Difference between the maximum and minimum weights & This shows the overall timespan of the group.
  \\\midrule

  \textbf{6-15} & Histogram of all weights & The histogram shows the distribution of time difference from the incoming stack trace to all stack traces in the group. In our experiments, we used the number of bins equal to 10. Each unique feature constitutes the value of the histogram in the corresponding bin. 
  \\\midrule\midrule
  
  \multicolumn{3}{c}{\textbf{Features based on both similarity and timestamps}}\\\midrule\midrule
  
  \textbf{16} & Weight of the first maximum & This shows how long ago the most similar stack trace was added to the group relative to the incoming stack trace, which can also affect how relevant it is.
  \\ \midrule
  
  \textbf{17-28} & Weighted similarity histogram & The weighted similarity histogram allows to estimate the distribution of similarities to the stack traces in the group, taking into account the time of their occurrence relative to the incoming stack trace. In our experiments, we used the number of bins equal to 12. Each unique feature constitutes the value of the histogram in the corresponding bin. 
  \\ \bottomrule
\end{tabular}
\label{table:features-description}
\end{table*}

After choosing a particular similarity model, we can calculate the values of the similarity between the incoming stack trace and all stack traces in all groups. Based on these similarities, we construct various features, the full list is presented in \Cref{table:features-description}.

\subsubsection{Features based on similarity}
The key feature that is used in the majority of works is the similarity value of the most similar stack trace in the group (we will refer to this feature as the \textit{first maximum}). 

\subsubsection{Features based on timestamps}

The idea behind using timestamps of stack traces lies in the assumption that not all stack traces in a group represent it the same. The groups tend to evolve over time, and, as we showed in Section~\ref{sec:motivation}, can include stack traces that are very different. For this reason, we believe that \textit{newer}, more recent stack traces can characterize the group better.
The information about the time of occurrence of stack traces will give our model the ability to independently decide which similarity value characterizes the similarity to the group the best.

Consider two stack traces: the $S_q$ stack trace (the incoming query stack trace that needs to be assigned to one of the groups) and the $S_g$ stack trace (the stack trace that is in one of the groups). Let us designate the time of their occurrence (timestamp) as $T_q$ and $T_g$, respectively.
For $\mathrm{similarity}(S_q, S_g)$, we define the weight $\mathrm{w}(S_q, S_g)$ as follows:
\begin{gather}
    \label{eq:weight}
    \mathrm{w}(S_q, S_g) = \frac{1}{\log(|T_q - T_g| + 1) + 1}.
\end{gather}

The idea behind this scale is that the closer in time two stack traces are to each other, the greater the weight is, with the values of the weight laying between $0$ and $1$. This allows prioritizing the similarity of stack traces that are closest in time to the incoming stack trace.

The following features can be constructed based on the proposed weights. The \textit{maximum weight} shows how long ago the group was updated relative to the incoming stack trace, the \textit{minimum weight} shows when the group was formed, and their \textit{difference} shows the timespan of the group. The \textit{mean weight} shows how far the given stack trace is from the group in time on average. The \textit{histogram of all weights} will also indicate the overall relation in time between the group and the incoming stack trace. To transform a histogram into a set of features, one simply needs to take the value of the histogram in each bin as a separate feature. Based on our preliminary experiments, we used a histogram with 10 bins.

\subsubsection{Features based on both similarity and timestamps}

Combining both the similarity and the temporal aspect, firstly, we can use the \textit{weight of the first maximum}~--- the weight corresponding to the \textit{first maximum}. This will allow us to understand how far the most similar stack trace is located in time from the incoming one.

Using the intuition presented in~\Cref{sec:motivation}, we would also like to have an idea about the distribution of the similarity values to all stack traces in the group.
This will make it possible to understand how widely the similarity values are scattered, which values prevail, and how much the value of the \textit{first maximum} differs from all others. For that, a similarity histogram can be employed (see \Cref{fig:hists}).

However, to take into account more information, instead of using the similarity histogram itself, we can use its modification that employs the proposed weights. To build a \textit{weighted similarity histogram}, it is necessary to assign a weight to each similarity value, which will allow us to move from the frequencies of occurrence of each similarity to their weighted frequencies. This means that each similarity value is included in the histogram not with a weight of 1, but with a weight calculated using Equation~\ref{eq:weight}. Based on our preliminary experiments, we used a histogram with 12 bins. 

Overall, this results in 28 features, the full list of which is presented in \Cref{table:features-description}. The selected features describe the relationships between stack traces and groups very fully, both in terms of similarity and time. We apply standard scaling to all the features before processing them.

\subsection{Linear Aggregation Model}

After all the described features are built, the only remaining thing is to transfer them to an \ag, rank the obtained similarities to groups, and assign the stack trace to the most similar group. We have tried several approaches, and in the result, selected a simple linear \ag, which is a weighted sum of constructed features. The practical advantages of such a model are that it is easy and fast to train, and is easy to introduce into production. To train the \ag, we used a RankNet Loss~\cite{ranknet}, and the feature weights were updated using the Adam optimizer~\cite{adam}.
\section{Evaluation}\label{sec:evaluation}

To test the applicability of our approach in the real world, we evaluated it not only on the open-source data, but also collected a dataset from JetBrains, a large software engineering company.
Specifically, we formulated three research questions:

\begin{questions}[leftmargin=1.3cm]
    \item Does the proposed \ag improve the performance of existing similarity models?
    \item Does the proposed \ag work better than the $k$-NN-based approach?
    \item Which features are the most important for the performance of the \ag?
\end{questions}

In this section, we describe how the experiments were set up, what data and metrics were used, as well as discuss the results of our experiments.

\subsection{Experimental setup}

Our goal is to compare several baseline similarity models from the literature with the \ag that employs them to consider all stack traces in the group instead of just the nearest one.
We evaluated five different similarity models described before: Modani et al.~\cite{modani}, Lerch and Mezini~\cite{lerch}, DURFEX~\cite{durfex}, TraceSim~\cite{tracesim}, and S3M~\cite{s3m}.
Additionally, we used CrashGraphs~\cite{crash_graphs} as a baseline, however, it cannot be aggregated since it does not calculate the similarity between individual stack traces.

It is necessary for all the models to see the same amount of information during training.
The data has a temporal component, used to construct some features, so it is very important for the train set to be located before the test set in time.

\begin{figure}[htbp]
\centering
    \includegraphics[width=\columnwidth]{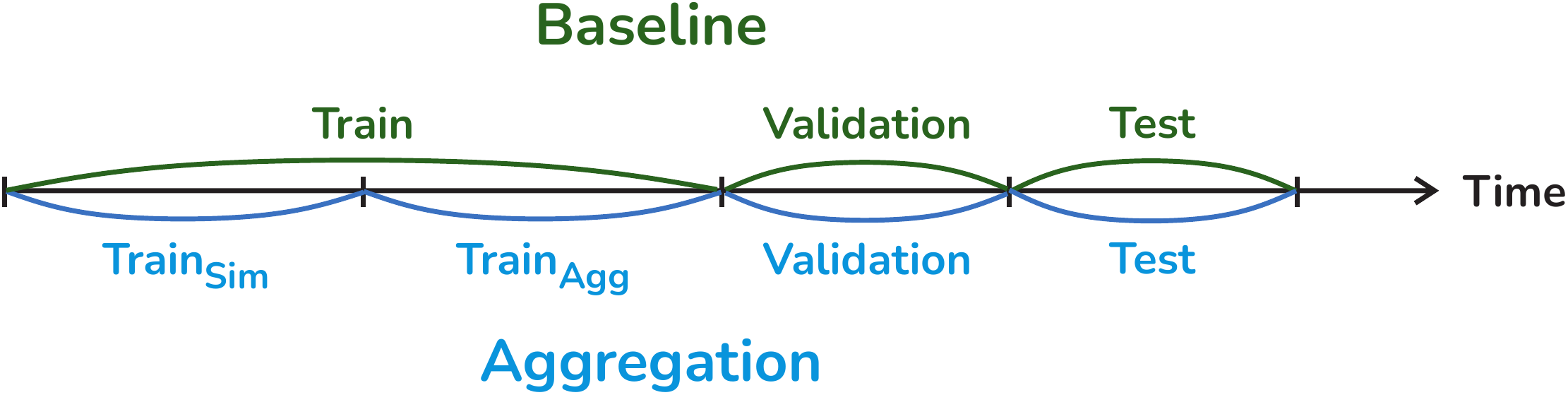}
    \centering
    \vspace{-0.4cm}
    \caption{The way the data was split.}
    \label{fig:data}
\end{figure}

The proposed way of splitting the data is shown in~\Cref{fig:data}.
All the models are validated on the $\mathrm{\textbf{Validation}}$ data span and tested on the $\mathrm{\textbf{Test}}$ data span, which contains the latest stack traces.
Each baseline similarity model is trained on the older data $\mathrm{\textbf{Train}}$.
The same data is then split into two intervals~--- $\mathrm{\textbf{Train}_\textbf{Sim}}$ and $\mathrm{\textbf{Train}_\textbf{Agg}}$.
On the $\mathrm{\textbf{Train}_\textbf{Sim}}$, the same similarity model is trained, with the help of which the similarity values on $\mathrm{\textbf{Train}_\textbf{Agg}}$ are obtained for training the \ag.
This way, the comparison is fair, the training for the \ag is merely divided into training the employed similarity model and training the aggregation itself.

\subsection{Data}

We carried out our experiments on two different datasets.
The first one is an open-source NetBeans dataset from our previous work~\cite{s3m}.
At the same time, we build our approach to be used in production, so we collected a new dataset from the proprietary data of JetBrains, a large software engineering company.
We plan to release this dataset upon acceptance to facilitate the further development of new approaches. 

\begin{table}[h]
\centering
\caption{The details of splitting the data into train, \\ validation (\textit{Val.}), and test sets.}
\begin{tabular}{ccccccc}
\toprule
\multirow{2}{*}{\textbf{Measure}} & 
\multicolumn{3}{c}{\textbf{NetBeans}} & \multicolumn{3}{c}{\textbf{JetBrains}}
\\ \cmidrule(lr){2-4}\cmidrule(lr){5-7} 
& \multicolumn{1}{c}{Train} & \multicolumn{1}{c}{Val.} & \multicolumn{1}{c}{Test} & \multicolumn{1}{c}{Train} & \multicolumn{1}{c}{Val.} & \multicolumn{1}{c}{Test} 

\\ \midrule

Groups &
30,975 & 1,695 & 6,084 & 6,356 & 1,077 & 1,731                          
\\ 

Reports & 
39,417 & 1,973 & 7,792 & 205,284 & 5,243 & 25,647                         
\\ 

Days & 
4,200 & 140 & 700 & 124 & 7 & 28                          
\\ \bottomrule
\end{tabular}

\label{tab:datasets}

\end{table}

\Cref{tab:datasets} shows the breakdown of datasets into train, validation, and test sets.
We split $\mathrm{\textbf{Train}}$ into $\mathrm{\textbf{Train}_\textbf{Sim}}$ and $\mathrm{\textbf{Train}_\textbf{Agg}}$ at the ratio of 20:1 for the NetBeans dataset and 9:1 for the JetBrains dataset.
This table also shows a strong difference between the datasets.
The NetBeans dataset generated an average of $7.7$ groups per day while the JetBrains dataset generated an average of $57.6$ groups per day.
Each group in the NetBeans dataset contains an average of $1.3$ reports, while for JetBrains, it is $25.8$ reports.
In short, the JetBrains data is less \textit{sparse} than the NetBeans data.
This diversity makes our data valuable for further research. Let us now describe each dataset a bit more.

\subsubsection{NetBeans dataset}
This dataset is open-source and was collected in our previous work~\cite{s3m}. NetBeans is an integrated development environment written in Java, the dataset includes their crash reports submitted before $2016$. Originally, the users attached files with a description of the failures that occurred, so the approach proposed by Lerch and Mezini~\cite{lerch} was used to generate crash reports by extracting stack traces from the attached files and description field using regular expressions.

\subsubsection{JetBrains dataset}\label{sec:jb_dataset}
JetBrains has its own system for handling automatically generated error reports. This system collects error reports from various projects written in JVM languages (Java, Kotlin, Scala), which helps to unify the work with them due to the homogeneous nature of the data. When a new report arrives, it is pre-processed. The following information is extracted from it: the product version, the point in time when the error occurred, the information about the system, and the stack trace itself.

New groups are formed as follows. When a new stack trace arrives, the Lerch and Mezini's similarity model~\cite{lerch} determines the stack trace's similarity to each of the groups and assigns the stack trace to the corresponding group. If the maximum similarity value is less than a certain threshold, then the corresponding stack trace is shown to the company developer who is experienced in this software component, and they themselves make a decision whether to define this stack trace in one of the existing groups or create a new one. To make a decision, the developer is shown a ranked list of groups.
Thus, data labeling is automatically performed whenever the developer has manually formed a new group or determined the stack trace into one of the existing groups. When working with our data, we do it in a similar way.

However, the usage of complex models in real-life large-scale projects introduces certain difficulties. Since the amount of data is really large, calculating all similarity values becomes very difficult, unacceptably long for production. For this reason, we use an additional \textit{filtration}. It works as follows.

Using the Lerch and Mezini's similarity model~\cite{lerch}, we find similarity values between the incoming stack trace and all stack traces in all groups. This model is very straightforward and fast, and can work the quickest. Then, having sorted the obtained values of similarity, we begin to select most similar stack traces until there are \textit{N} unique groups. This approach allows us to reduce the number of groups and reports used when inferencing the models. In our experiments, we used $N = 1000$. When choosing the value of \textit{N}, we were guided by the performance of the entire system (the smaller the value of \textit{N}, the faster the inference of the models goes) and the value of the $\mathrm{RR@N}$ metric, which is obtained using the similarity model. ``Heavier'' similarity models, for example, deep learning-based ones, can then be used on this filtered subset of data much faster. Evaluating models with filtering is crucial for their use in industrial systems.

Finally, it should be noted that the company's system has the following important property. If a group has not been updated for more than two months ($62$ days), then it is automatically considered closed for new error reports. That is, groups that have not been updated in the last $62$ days are not considered for a newly arrived stack trace. Thus, this feature must be taken into account when carrying out experiments.

\subsection{Performance Metrics}

\begin{table*}[t]
\centering
\caption{The results of the experiments of different similarity models with and without the aggregation. \textbf{Base} indicates the metric value for the given baseline similarity model, \textbf{Agg} indicates the result with the \ag, \textbf{$\Delta_{A}$} indicates the increase that the aggregation brings to the baseline, \textbf{$k$-NN} indicates the result with the $k$-NN aggregation, \textbf{$\Delta_{K}$} indicates the increase that the $k$-NN brings to the baseline.}
\begin{tabular}{@{}ccccccccccccc@{}}
\toprule
\multirow{2}{*}{\textbf{Dataset}} & 
\multirow{2}{*}{\textbf{Model}} & 
\multirow{2}{*}{\begin{tabular}{@{}c@{}}\textbf{Best $k$-NN} \\ \textbf{Configuration}\end{tabular}} &
\multicolumn{5}{c}{\textbf{MRR}} & 
\multicolumn{5}{c}{\textbf{RR@1}} 
\\ 
\cmidrule(lr){4-8}\cmidrule(lr){9-13} & & &
\textbf{Base} & \textbf{Agg} & \textbf{$\Delta_{A}$} & \textbf{$k$-NN} & \textbf{$\Delta_{K}$} &
\textbf{Base} & \textbf{Agg} & \textbf{$\Delta_{A}$} & \textbf{$k$-NN} & \textbf{$\Delta_{K}$}         
\\ 
\midrule
\multirow{6}{*}{\textbf{NetBeans}} 
& CrashGraphs~\cite{crash_graphs} & — & 
0.33 & — & — & — & — &  
0.25 & — & — & — & —
\\ 
& Modani et al.~\cite{modani} & Triweight, $k = 5$ &  
0.36 & 0.48 & +\,0.08 & 0.39 & \textbf{+\,0.03} &  
0.28 & 0.36 & +\,0.08 & 0.30 & +\,0.02
\\
& Lerch and Mezini~\cite{lerch} &  Triweight, $k = 3$ & 
0.37 & 0.49 & \textbf{+\,0.12} & 0.38 & +\,0.01 & 
0.26 & 0.41 & \textbf{+\,0.15} & 0.28 & +\,0.02
\\
& DURFEX~\cite{durfex} & Triweight, $k = 13$ & 
0.42 & 0.50 & +\,0.08 & 0.45 & \textbf{+\,0.03} &
0.32 & 0.42 & +\,0.10 & 0.35 & +\,0.03
\\
& TraceSim~\cite{tracesim} & Triweight, $k = 7$ & 
0.39 & 0.48 & +\,0.09 & 0.41 & +\,0.02 &
0.30 & 0.38 & +\,0.08 & 0.32 & +\,0.02
\\
& S3M~\cite{s3m} &  Quartic, $k = 13$ & 
\textbf{0.46} & \textbf{0.56} & +\,0.10 & \textbf{0.49} & \textbf{+\,0.03} &
\textbf{0.35} & \textbf{0.47} & +\,0.12 & \textbf{0.39} & \textbf{+\,0.04}
\\ 
\midrule
\multirow{6}{*}{\textbf{JetBrains}} 
& CrashGraphs~\cite{crash_graphs} & — &  
0.73 & — & — & — & — &
0.64 & — & — & — & —
\\
& Modani et al.~\cite{modani} &  Triweight, $k = 3$ & 
0.80 & 0.81 & +\,0.01 & 0.80 & 0 &
0.72 & 0.75 & +\,0.03 & 0.73 & +\,0.01
\\
& Lerch and Mezini~\cite{lerch} &  Triweight, $k = 3$ & 
0.78 & 0.83 & \textbf{+\,0.05} & 0.79 & \textbf{+\,0.01} &
0.68 & 0.76 & \textbf{+\,0.08} & 0.69 & +\,0.01
\\
& DURFEX~\cite{durfex} &  Triangle, $k = 3$ & 
0.81 & 0.84 & +\,0.03 & 0.82 & \textbf{+\,0.01} &
0.73 & 0.78 & +\,0.05 & 0.74 & +\,0.01
\\
& TraceSim~\cite{tracesim} &  Triweight, $k = 3$ & 
0.80 & 0.81 & +\,0.01 &  0.81 & \textbf{+\,0.01} &
0.72 & 0.74 & +\,0.02 & 0.74 & \textbf{+\,0.02}
\\
& S3M~\cite{s3m} &  Uniform, $k = 3$ & 
\textbf{0.87} & \textbf{0.91} & +\,0.04 & \textbf{0.88} & \textbf{+\,0.01} &
\textbf{0.80} & \textbf{0.86} & +\,0.06 & \textbf{0.81} & +\,0.01
\\ 
\midrule\midrule
\multirow{2}{*}{\textbf{Dataset}} & 
\multirow{2}{*}{\textbf{Model}} & 
\multirow{2}{*}{\begin{tabular}{@{}c@{}}\textbf{Best $k$-NN} \\ \textbf{Configuration}\end{tabular}} &
\multicolumn{5}{c}{\textbf{RR@5}} & 
\multicolumn{5}{c}{\textbf{RR@10}} 
\\ 
\cmidrule(lr){4-8}\cmidrule(lr){9-13} & & &
\textbf{Base} & \textbf{Agg} & \textbf{$\Delta_{A}$} & \textbf{$k$-NN} & \textbf{$\Delta_{K}$} &
\textbf{Base} & \textbf{Agg} & \textbf{$\Delta_{A}$} & \textbf{$k$-NN} & \textbf{$\Delta_{K}$}         
\\ 
\midrule
\multirow{6}{*}{\textbf{NetBeans}} 
& CrashGraphs~\cite{crash_graphs} &  — & 
0.43 & — & — & — & — &  
0.50 & — & — & — & —
\\ 
& Modani et al.~\cite{modani} &  Triweight, $k = 5$ & 
0.46 & 0.57 & \textbf{+\,0.11} & 0.49 & \textbf{+\,0.03} &  
0.51 & 0.64 & \textbf{+\,0.13} & 0.54 & \textbf{+\,0.03}
\\
& Lerch and Mezini~\cite{lerch} &  Triweight, $k = 3$ & 
0.49 & 0.58 & +\,0.09 & 0.51 & +\,0.02 & 
0.57 & 0.65 & +\,0.08 & 0.58 & +\,0.01
\\
& DURFEX~\cite{durfex} & Triweight, $k = 13$ & 
0.56 & 0.60 & +\,0.04 & 0.58 & +\,0.02 &
0.61 & 0.65 & +\,0.04 & 0.62 & +\,0.01
\\
& TraceSim~\cite{tracesim} & Triweight, $k = 7$ & 
0.50 & 0.60 & +\,0.10 & 0.52 & +\,0.02 &
0.55 & 0.67 & +\,0.12 & 0.57 & +\,0.02
\\
& S3M~\cite{s3m} &  Quartic, $k = 13$ & 
\textbf{0.60} & \textbf{0.68} & +\,0.08 & \textbf{0.61} & +\,0.01 &
\textbf{0.66} & \textbf{0.73} & +\,0.07 & \textbf{0.67} & +\,0.01
\\ 
\midrule
\multirow{6}{*}{\textbf{JetBrains}} 
& CrashGraphs~\cite{crash_graphs} &  — & 
0.83 & — & — & — & — &
0.86 & — & — & — & —
\\
& Modani et al.~\cite{modani} &  Triweight, $k = 3$ & 
0.89 & 0.89 & 0 & 0.89 & 0 &
0.91 & 0.91 & 0 & 0.91 & 0
\\
& Lerch and Mezini~\cite{lerch} &  Triweight, $k = 3$ & 
0.90 & 0.91 & +\,0.01 & 0.90 & 0 &
0.92 & 0.93 & \textbf{+\,0.01} & 0.92 & 0
\\
& DURFEX~\cite{durfex} &  Triangle, $k = 3$ & 
0.90 & 0.92 & \textbf{+\,0.02} & 0.90 & 0 &
0.93 & 0.94 & \textbf{+\,0.01} & 0.93 & 0
\\
& TraceSim~\cite{tracesim} &  Triweight, $k = 3$ & 
0.89 & 0.89 & 0 & 0.89 & 0 &
0.91 & 0.92 & \textbf{+\,0.01} & 0.91 & 0
\\
& S3M~\cite{s3m} &  Uniform, $k = 3$ & 
\textbf{0.96} & \textbf{0.97} & +\,0.01 & \textbf{0.97} & \textbf{+\,0.01} &
\textbf{0.97} & \textbf{0.97} & 0 & \textbf{0.97} & 0
\\ 
\bottomrule
\end{tabular}
\vspace{-0.2cm}
\label{tab:results}
\end{table*}

We used the Mean Reciprocal Rank ($\mathrm{MRR}$)~\cite{mrr} and the Recall Rate ($\mathrm{RR@k}$)~\cite{rr} as comparison metrics. The choice of these metrics is based on research from previous works~\cite{lerch,durfex,irving}. The Mean Reciprocal Rank (MRR) reflects the ranking quality of the entire list, as it calculates the average reciprocal position of the correct item:

\begin{gather}
    \mathrm{MRR} = \frac{1}{|Q|}\sum\limits_{i=1}^{|Q|}\frac{1}{\mathrm{rank}_i}.
\end{gather}

That is, for each incoming stack trace $\in Q$, we get a ranked list of groups and find the position of the only correct answer in it.
If the stack trace was the first in the group, then there will be no correct answer in the list and its position can be considered equal to infinity.
After that, we take the average over all queries from the reciprocal values of the positions of the correct elements.
In the absence of a valid element, the inverse value will be zero.

Another important metric, \textit{Recall Rate at the first $k$ positions} ($\mathrm{RR} @ k$) counts the proportion of cases when the correct group was among the first $k$ options:

\begin{gather}
    \mathrm{RR}@k = \frac{1}{|Q|}\sum\limits_{i=1}^{|Q|}[\mathrm{rank}_i \leq k].
\end{gather}

This metric is more interpretable, which helps to better predict the future behavior of the system.
However, it only counts the first $k$ elements of the list, ignoring the rest.
The $\mathrm{RR} @ 1$ metric is especially important for us, since the automatic report processing system selects exactly one closest group and places the report there.
Accordingly, for such a system, the main quality criterion will be precisely the accuracy in the first position, since all the others will not be taken into account.
In addition, we are interested in the $\mathrm{RR} @ 5$ and $\mathrm{RR} @ 10$ metrics, since in the case of manual group selection, the developer is shown the top relevant groups.

\subsection{Training}

Since we are solving the ranking problem, we used RankNet Loss~\cite{ranknet} to train the \ag. For every positive example, we pick ten random negative examples as a negative sampling, which allows the model to better distinguish between similar stack traces.
Initially, the feature coefficients are randomly initialized. We used the Adam optimizer~\cite{adam} with the learning rate of $1\mathrm{e}{-3}$ and the weight decay of $1\mathrm{e}{-3}$.

\subsection{$k$-NN-based Approach}

To test the usefulness of the proposed \ag and make sure that it is necessary, we also implemented an alternative, simpler version of aggregating the data from different stack traces in the group.
We use the $k$-NN algorithm described in~\Cref{sec:knn} to take into account the distances to all the stack traces in all groups.

Since most approaches based on the idea of $k$-NN use distance values between objects, it is first necessary to transform the obtained similarity values in such a way that they satisfy the following property: the larger the similarity value, the smaller the distance value.
Thus, the more similar stack traces are, the closer they will be located to each other.
To comply with the required property, we used the following transformation.
Consider the similarity values $\{s_i\}_{i=1}^{M}$ from a given stack trace to all stack traces from all groups.
Then, the new values of the distances will be determined using the following formula:
\begin{gather}
    d_i = \max_{j=1,\ldots,M}\{s_j\} - s_i,\ \forall i \in 1,\ldots,M. 
    \label{distance_from_similarity}
\end{gather}

Some $k$-NN approaches use a support from 0 to 1, that is, if the distance is greater than or equal to 1, then the weight of the object will be zero. To take this into account, all distances obtained through Equation~\ref{distance_from_similarity} are divided by the distance to ($k$~+~1)-th neighbor.

Note that the distances obtained by Equation~\ref{distance_from_similarity} satisfy the requirement: the distance to the most similar stack trace will be minimal, and the less similar the stack traces are, the greater the resulting distance value will be.

In our experiments, we used weighted $k$-NN and evaluated the following eleven kernel functions: uniform, triangle, epanechnikov, quartic, triweight, gaussian, cosine, tricube, logistic, sigmoid, and silverman, as well as the apporaches proposed by Hyukjun et al.~\cite{k_conditional} and Ekin et al.~\cite{distance_based} For both datasets and all the tested similarity models, we tested these weighting methods and values of $k$ from 1 to 15, since larger values of $k$ always demonstrated worse performance. 

\subsection{Results}

\begin{figure*}[htbp]
\centering
    \includegraphics[width=\textwidth]{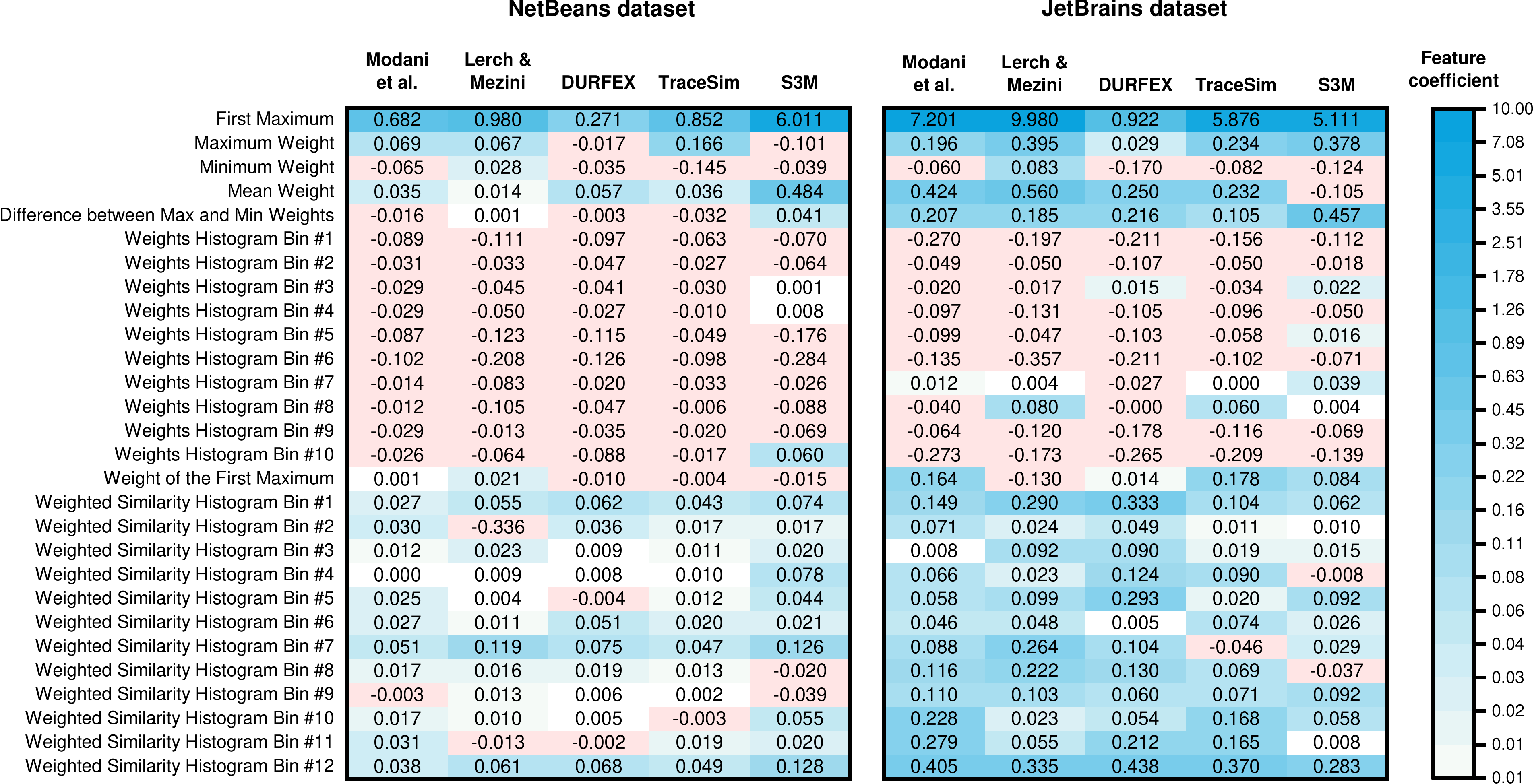}
    \centering
    \vspace{-0.4cm}
    \caption{The heatmap of feature importance for various models on the both datasets.}
    \label{fig:jb_features}
\end{figure*}

Let us now describe the results of our experiments and answer the posed research questions.

\subsubsection{RQ1: Aggregation Model vs. Baselines}

In the first research question, we wanted to see how the proposed \ag improves the performance of different similarity models. The results of the experiments are shown in~\Cref{tab:results}. For all four metrics (MRR, $\mathrm{RR} @ 1$, $\mathrm{RR} @ 5$, $\mathrm{RR} @ 10$), the column $\Delta_{A}$ shows the improvement of the \ag over the given baseline model.
Our \ag showed the largest improvement for the approach of Lerch and Mezini~\cite{lerch}: using the \ag improves the quality metric $\mathrm{RR} @ 1$ by $15$ percentage points on the NetBeans dataset and by $8$ percentage points on the JetBrains dataset. At the same time, the highest value in the $\mathrm{RR} @ 1$ metric overall was obtained with the aggregation of similarities obtained using S3M ($0.47$ and $0.86$ on the NetBeans dataset and the JetBrains dataset, respectively). Overall, it can be seen that using the \ag allows us to significantly improve the performance of all the models.

It can also be seen that the improvement is smaller on the new dataset than on the open-source NetBeans dataset. This difference in the influence demonstrates its importance for further research: improving the aggregation or developing brand new methods both require representative industrial data.

Finally, it should be mentioned that the filtration described in Section~\ref{sec:jb_dataset} greatly speeds up the processing of large amounts of data in the real run in production. We tested the approach on a machine with 8 Intel Xeon CPUs @ 2.30 GHz and 60 GB of RAM, and the report processing speed increased up to $200$ times.

\observation{According to our results, the proposed Aggreation Model increases the quality of the considered similarity models up to $8$ percentage points on the JetBrains dataset and up to $15$ percentage points on the open-source NetBeans dataset. The approach is very simple to implement in a real-world project and can significantly increase the quality of the employed models.}

\subsubsection{RQ2: Aggregation Model vs. $k$-NN}

In the second research question, we studied how the performance of our \ag compares with $k$-NN based approaches.
The results of applying $k$-NN-based approaches are also presented in \Cref{tab:results}. For each similarity model, the table shows the $k$-NN configuration that showed the best results (the kernel function and $k$).
The column $\Delta_{K}$ shows the improvement of the best $k$-NN model over the given baseline, thus, to compare our \ag and the $k$-NN-based approach, one needs to compare columns $\Delta_{A}$ and $\Delta_{K}$. 

The increase in the $\mathrm{RR} @ 1$ metric when using $k$-NN-based approaches is not as significant as in the case of the \ag. 
On the open-source NetBeans data, the maximum increase in the $\mathrm{RR} @ 1$ metric turned out to be $4$ percentage points for the S3M~\cite{s3m} model, while in the case of the \ag, the maximum increase in the $\mathrm{RR} @ 1$ metric turned out to be $15$ percentage points for the Lerch and Mezini~\cite{lerch} model.
As for the JetBrains data, the $k$-NN-based approaches demonstrated even worse results.
The maximum increase in the $\mathrm{RR} @ 1$ metric turned out to be $2$ percentage points for the TraceSim~\cite{tracesim} model, while in the case of the \ag, the maximum increase in the $\mathrm{RR}@1$ metric was $8$ percentage points for the Lerch and Mezini~\cite{lerch} model.
Overall, the best-performing $k$-NN-based models never outperformed the \ag.

\observation{The results of applying the $k$-NN-based approach are inferior in quality to the \ag for all the similarity models except for TraceSim.
This indicates the usefulness of the aggregation and the importance of taking into account the temporal aspect of stack traces.}

\subsubsection{RQ3: Feature Importance}
Finally, in the third research question, we assessed the importance of the features used in the \ag. Since we used standard scaling, and all the features have the same scale, we can analyze the coefficients of the used linear model. \Cref{fig:jb_features} shows the coefficients of each of the features listed in~\Cref{table:features-description} for each similarity model, trained for both datasets.

First of all, it can be seen that for all similarity models, the feature \textit{first maximum} (\textit{i.e.}, the similarity to the most similar stack trace) has the largest coefficient on both datasets. This confirms the importance of its use and explains the good results of baseline similarity models.

It can also be seen that for all similarity models, the following is true: the feature \textit{weighted similarity histogram bin \#12} has the largest coefficient out of all the bins of the weighted similarity histogram on the JetBrains dataset.
This is also true for almost all similarity models on the open-source NetBeans data.
Bin \#12 contains the most similar stack traces, including the \textit{first maximum}, and the more similar the stack traces are, the greater the score the \ag gives to this group.

Finally, it can be seen that most of the coefficients of the \textit{weights histogram} features have negative values. 
This is consistent with our assumption that the more stack traces there are that are far removed in time, and the longer the group has not been updated, the less likely it is that the stack trace should be assigned to this group. 
Thus, the \ag introduces some penalty to the final similarity value.

\vspace{0.1cm}

\observation{Expectedly, the \ag gives the greatest preference to the \textit{first maximum} feature, which is the feature that is used in virtually all existing methods.
However, the increase in performance and the analysis show that the model uses all the information given to it.}

\vspace{0.1cm}
\section{Threats to Validity}\label{sec:threats}
In our study, the following threats to validity can be found:
\begin{itemize}
    \item \textbf{Subject selection bias.}
    The performance of machine learning algorithms depends on the data on which the algorithms are trained and on which they are applied.
    In the case of our approach, we have two machine learning algorithms that are used to solve the deduplication problem: the similarity model and the \ag itself, which uses the calculated similarity values.
    To mitigate this threat, we tested our approach on two datasets: open-source NetBeans data and the new data from the JetBrains software company.
    The difference in results highlights the importance of sharing new data with the community.
    
    \item \textbf{Limited scope of application.}
    Our approach requires a sufficient number of crash reports containing the information about stack traces to train the similarity model and the \ag, while also having information about their time of occurrence.
    However, this allows our approach to be useful for any moderately large industrial system.
    It is also possible to experiment with using a pre-trained model in a new project.
    
    \item \textbf{Programming language bias.}
    We evaluated our approach on two stack trace datasets that were both collected for JVM languages.
    For stack traces in other programming languages, both the importance of the features used and the results of applying the \ag may differ, because the result of its work directly depends on the similarity values obtained from the used similarity methods.
    Further research is needed to assess how well the proposed approach generalizes to other languages and systems.
\end{itemize}
\section{Conclusion}\label{sec:conclusion}
In this paper, we described a new approach for solving the problem of grouping stack traces that uses not only the information about the values of similarity between stack traces, but also the information about their time of occurrence.
We evaluated our approach on the open-source NetBeans dataset, but also collected a new dataset from the proprietary data of JetBrains, a large software engineering company.
The implementation of our approach is available online on GitHub: \url{https://github.com/nkarasovd/AggregationModel}.
Upon acceptance, we plan to publish our dataset to facilitate further research and its industrial application.

Our experiments have demonstrated the superiority of our approach over the state-of-the-art methods by $15$ and $8$ percentage points in Recall Rate Top-$1$ metric on both the open-source NetBeans data and our dataset, respectively.
Also, using simpler $k$-NN-based approaches did not allow us to obtain the same increase in performance. 

Our work can be continued in different directions.
Firstly, we plan to improve the \ag itself, experimenting with different features.
Secondly, we want to experiment with a neural network architecture that would create embeddings of stack traces. This will allow us to create an aggregated representation of a group using the embeddings of stack traces in it and the incoming stack trace.
Finally, it is of great interest to improve the results of different models for real-world applications.
In this regard, the provided JetBrains dataset will be of particular use, since both types of aggregation demonstrated smaller improvements on such data.

\bibliographystyle{IEEEtran}
\balance
\bibliography{IEEEabrv,paper}

\end{document}